# Strain-Induced Decoupling Drives Gold-Assisted Exfoliation of Large-Area Monolayer 2D Crystals

*Jakob Ziewer[†], Abyay Ghosh[†], Michaela Hanušová[‡§], Luka Pirker[‡], Otakar Frank[‡], Matěj Velický[‡], Myrta Grüning[†], and Fumin Huang[†]*

[†] Centre for Quantum Materials and Technologies, School of Mathematics and Physics, Queen's University Belfast, University Road, Belfast, BT7 1NN, United Kingdom

[‡] J. Heyrovský Institute of Physical Chemistry, Czech Academy of Sciences, Dolejškova 2155/3, Prague, 18223, Czech Republic

[§] Faculty of Chemical Engineering, University of Chemistry and Technology, Prague, Technická 5, 166 28 Prague 6, Czech Republic

ABSTRACT: Gold-assisted exfoliation (GAE) is a groundbreaking mechanical exfoliation technique, producing centimeter-scale single-crystal monolayers of two-dimensional (2D) materials. Such large, high-quality films offer unparalleled advantages over the micron-sized flakes typically produced by conventional exfoliation techniques, significantly accelerating the research and technological advancements in the field of 2D materials. Despite its wide applications, the fundamental mechanism of GAE remains poorly understood. In this study, using $MoS_2$ on Au as a model system, we employ ultra-low frequency Raman spectroscopy to elucidate how the interlayer interactions within $MoS_2$ crystals are impacted by the gold substrate. The results reveal that the coupling at the first $MoS_2$-$MoS_2$ interface between the adhered layer on the gold substrate and the adjacent layer, is substantially weakened, with the binding force being reduced to nearly zero. This renders the first interface the weakest point in the system, thereby the crystal preferentially cleaves at this junction, generating large-area monolayers with sizes comparable to the parent crystal. Biaxial strain in the adhered layer, induced by the gold substrate, is identified as the driving factor for the decoupling effect. We establish the strain-induced decoupling effect as the primary mechanism of GAE, which could also play a significant role in general mechanical exfoliations.

KEYWORDS: 2D materials, $MoS_2$, gold-assisted exfoliation, strain, decoupling

INTRODUCTION

Gold-assisted exfoliation (GAE) is a groundbreaking mechanical exfoliation technique capable of producing centimeter-scale single-crystal monolayers of a diverse range of two-dimensional (2D) materials.[1-3] It expands the exfoliated film area by several orders of magnitude compared to the conventional technique using scotch tape and $SiO_2$ substrates.[4] Although bottom-up synthesis methods, such as physical vapor deposition (PVD) and chemical vapor deposition (CVD), can yield wafer-scale monolayers,[5–7] their quality is often compromised by defects and polycrystallinity.[8,9] By contrast, GAE produces pristine monolayer crystals of superior quality.[10] Combining scalability and excellent material quality, GAE provides unprecedented opportunities for advanced research and technological development in the field of 2D materials, becoming central to the construction of a wide variety of structures including free-standing monolayers, heterostructures, and moiré superlattices.[11-14]

The GAE technique is broadly applicable to 2D materials containing sulfur (S), selenium (Se), or tellurium (Te), such as transition metal dichalcogenides (TMDCs) including $MoS_2$, $WS_2$, $MoSe_2$, $WSe_2$, $MoTe_2$, $WTe_2$, and various other chalcogenides like InSe, GaSe, $In_2Se_3$, $Cr_2Ge_2Te_6$, and $Fe_3GeTe_2$.[1,2] It has also been applied to a few non-chalcogenide 2D materials, such as graphene, hexagonal boron nitride (hBN), black phosphorus, and $CrCl_3$.[2,12,15] To date, more than fifty layered materials have been successfully exfoliated with GAE.[2,16] Recently, the technique has been extended to alternative metals including Ag, Pd, Cu, Ni, and Co.[17-19]

Despite its extensive use and significant impacts in the field of 2D materials, the fundamental mechanism of GAE and, more generally, metal-assisted exfoliation, remains poorly understood. Adhesion between 2D crystals and the gold substrate has been proposed as a key factor.[1,2,20] It is assumed for the large-area exfoliation to take place, the adhesion of the 2D crystal to the substrate needs to exceed the interlayer van der Waals (vdW) force within the 2D crystal.[1,2] However, this fails to explain why GAE predominantly produces large-area monolayers with near-unity yield, instead of randomly generating flakes of various thickness as is the case for the scotch tape/SiO$_2$ method. Substrate-induced strain has been hypothesized as a potential driver weakening the coupling at the first interface between the adhered layer and the adjacent layer, favoring monolayer exfoliation.[3,20-22] However, these are only speculations. To date, solid experimental evidence supporting this hypothesis has been lacking, leaving critical aspects of the mechanism unresolved.

In this work, we present compelling experimental evidence to reveal the fundamental mechanism of GAE. Using MoS$_2$ on gold as a model system,[1-3] we characterize the force constants at the first MoS$_2$-MoS$_2$ interface, i.e. the interface between the bottommost layer adhered to the gold substrate and the adjacent upper layer, through ultra-low frequency (ULF) Raman spectroscopy.[23-26] Our findings reveal that the coupling at the first interface is significantly weakened during the exfoliation process, with the degree of decoupling depending on crystal thickness. For bilayers, the coupling is weakened by about 20%, which increases to ~50% for tetralayers, and nearly 100% for crystals thicker than five layers. These results indicate that the first interface is the weakest point in the system. Therefore, MoS$_2$ crystals preferentially cleave at this interface, generating large-area monolayers on the gold substrate. This is further supported by the observation of micron-sized bubbles in annealed samples. When the bubbles burst at elevated temperatures, the exposed surfaces are all monolayers, unambiguously confirming that the first interface is the weakest point in the system. Biaxial strain in the adhered layer, induced by the gold substrate, is identified to be the primary driver of the decoupling effect. Density functional theory (DFT) simulations reveal that when the strain in the adhered layer and the crystal thickness exceed specific thresholds, monolayer exfoliation becomes energetically favorable.

RESULTS AND DISCUSSIONS

**Sample Fabrication and Thickness Characterization**

The processes of GAE have been described in a previous study,[1] briefly summarized here: MoS$_2$ crystals are cleaved just prior to exfoliation to minimize contamination; the cleaved surface is then pressed onto a freshly prepared Au substrate and lifted off, resulting in large-area monolayers adhered to the Au substrate. A range of samples are prepared on thin Au films of various thickness (2, 5 and 10 nm) as well as on commercially purchased template-stripped (TS) Au film of 100 nm. All thin Au films (unless otherwise stated) are deposited on a SiO$_2$(100 nm)/Si substrate (referred to as SiO$_2$ hereafter) with a 1 nm Ti adhesion layer through magnetron sputtering. For all Au samples, the exfoliated MoS$_2$ crystals are predominantly monolayers exceeding 1 mm in lateral dimension (limited by parent crystal size), as shown in Figure 1a and Supplementary Figure S1. Besides the monolayers, there are usually some scarcely distributed multilayers (Figure 1b). For comparison, we also prepare MoS$_2$ crystals on bare SiO$_2$ substrates as reference samples, which comprise small flakes (<20 um) of various thicknesses (Figure 1c). More details of sample fabrication are described in Methods and in previous publications.[1,27]

The thickness of MoS$_2$ layers is determined through multiple characterizations, including optical contrast, Raman, and AFM measurements. For example, the intensity ratios between the Raman modes ($E_{2g}^1$ and $A_{1g}$ modes, using the bulk notation) of MoS$_2$ and the underlying Si substrate (520.7 cm$^{-1}$ peak) are shown to be quasi-linearly proportional to the layer number (Figure 1d, 1e), which can be used to accurately determine the thickness of MoS$_2$.[28] This, corroborated with optical contrast, ULF Raman, and AFM measurements (Figure S2), allows us to accurately identify 1-10 layer MoS$_2$ on Au and SiO$_2$.

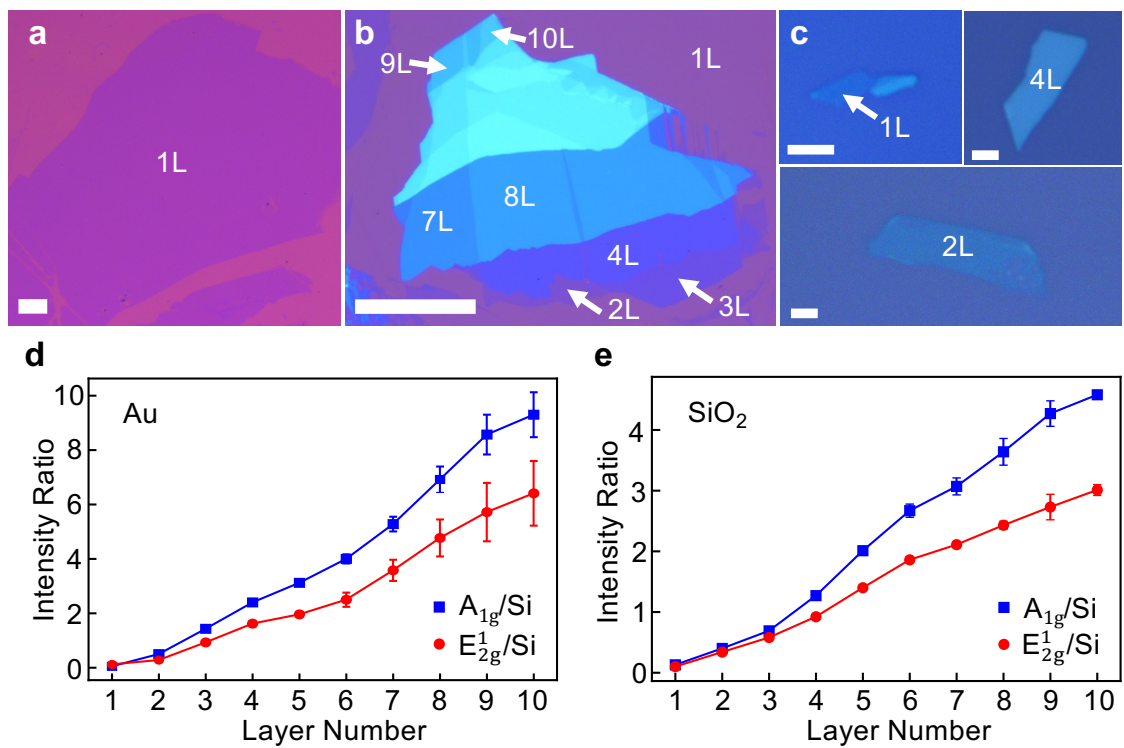

**Figure 1.** Optical images of MoS$_2$ and thickness characterization. Optical images of large-area monolayer (a) and few-layer (b) MoS$_2$ exfoliated on a 10 nm Au substrate. Scale bars: 100 $\mu$m. (c) Optical images of MoS$_2$ exfoliated on a SiO$_2$(100 nm)/Si substrate. All scale bars are 2 $\mu$m. (d-e) Raman intensity ratios of the MoS$_2$ E$_{2g}^1$ and A$_{1g}$ peaks, normalized to the intensity of the 520.7 cm$^{-1}$ Si peak for 1-10L MoS$_2$ on a 10 nm Au substrate (d) and a SiO$_2$(100 nm)/Si substrate (e).

**High-Frequency Raman Spectra: Strain in the Adhered Layer**

We first investigate the high-frequency Raman spectra of MoS$_2$. Figure 2 shows examples of the high-frequency Raman spectra of MoS$_2$ on 10 nm Au (Figure 2a) and SiO$_2$ (Figure 2b) substrates. The high-frequency Raman spectra show two main vibration modes, E$_{2g}^1$ mode (in-plane) and A$_{1g}$ mode (out-of-plane). The frequency and intensity of both modes are thickness dependent. For MoS$_2$ films on SiO$_2$, both the E$_{2g}^1$ mode and the A$_{1g}$ mode display a single peak. The A$_{1g}$ mode upshifts while the E$_{2g}^1$ mode downshifts with increasing thickness, in agreement with prior literature.[29]

For MoS$_2$ on Au, the spectra are notably different. The E$_{2g}^1$ mode of a monolayer displays a single peak at 380 cm$^{-1}$ (Figure 2d), which is downshifted by 6 cm$^{-1}$ with respect to its SiO$_2$ counterpart (386 cm$^{-1}$). The downshift of the E$_{2g}^1$ mode indicates the presence of a biaxial strain in the monolayer.[27,30] Based on the frequency shift, the mean strain is estimated to be 1.1% using a literature-averaged Grüneisen parameters of 0.71 (using the formula $\varepsilon = \frac{\Delta\omega}{2\gamma\omega_0}$, $\Delta\omega$ is frequency shift, $\omega_0$ is the frequency at zero strain, and $\gamma$ is the Grüneisen parameter),[30–32] close to the 1.2% reported previously.[27]

The E$_{2g}^1$ mode of 2-4L MoS$_2$ on Au shows splitting, which can be fitted with two Lorentzian peaks (Figure 2d). The peak positions of Raman modes averaged from several spectra are presented in Figure 2c. As seen from Figures 2(c-d), the low-frequency peaks of the MoS$_2$ E$_{2g}^1$ doublets on Au (denoted as E(L) in Figure 2c) remain around 380 cm$^{-1}$ for the bilayer, trilayer, and tetralayer, closely matching that of the monolayer. This suggests that the E(L) peak originates from the bottom strained layer adhered to the Au substrate. On the other hand, the high-frequency components of the E$_{2g}^1$ doublets match those on SiO$_2$, implying that they originate from the top layer(s), which are little affected by the strained bottom layer. This indicates that the biaxial strain is mostly localized in the bottom adhered layer with little transfers to subsequent layers, consistent with theoretical simulations.[22,33] The intensity of the E(L) peak decreases with thickness (Figure 2d), becoming hardly observable in crystals thicker than five layers. This is likely due to the combined effects from light blockage and absorption by the top layers as well as the signal being obscured by the enhanced Raman signals of the top layers, when the crystal thickness increases. These results are observed on all Au samples, with some variance in the shift of the monolayer E(L) peaks with the lowest being (377.8±0.1) cm$^{-1}$ for 2 nm Au samples and the highest being (380.0 ± 0.1) cm$^{-1}$ for 5 nm Au samples, indicating a strain ranging from roughly 1.5% to 1.1%.

Unlike the $E^1_{2g}$ mode, the $A_{1g}$ mode of monolayer MoS$_2$ on Au shows a distinct doublet, with one peak at 396.0 cm$^{-1}$ and the other at 404.5 cm$^{-1}$. Weak splitting is also observed for the $A_{1g}$ modes of few-layer MoS$_2$ on Au, as shown by the small humps (Figure 2d). The high frequency peak (404.5 cm$^{-1}$) closely matches that on SiO$_2$. The appearance of the 396.0 cm$^{-1}$ peak is attributed to doping effects,[27,34] suggesting a doped-electron concentration of $3.6 \times 10^{13}$ cm$^{-2}$ (equivalent of 0.033 extra electrons per unit cell).[35] The $E^1_{2g}$ mode is sensitive to strain but less to doping, while the $A_{1g}$ mode is sensitive to doping but not significantly to strain. For the strain and doping observed here the covariance of the $E^1_{2g}$ and $A_{1g}$ modes are negligible and shifts in each can be taken to arise solely from strain or doping respectively.[34]

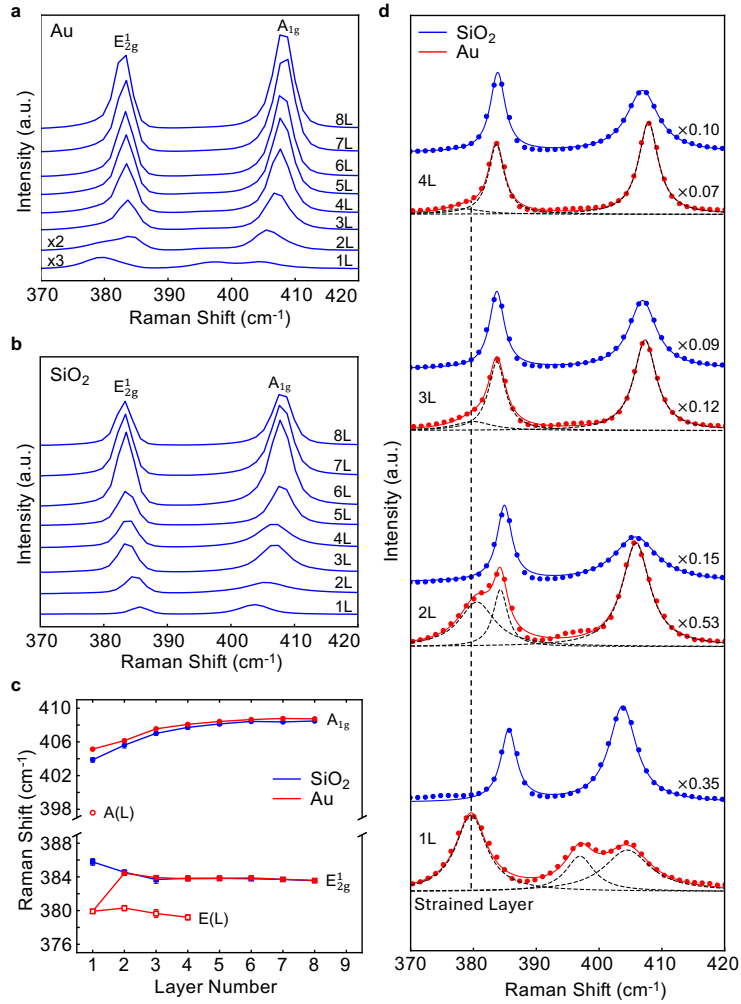

**Figure 2.** High-frequency Raman spectra of MoS$_2$. Raman spectra of 1-8L MoS$_2$ on 10 nm Au (a) and SiO$_2$ substrates (b). (c) Peak frequencies of the Raman modes. The $E_{2g}^1$ modes of 2-4L MoS$_2$ on Au consist of two peaks, a low-frequency peak E(L) around 380 cm$^{-1}$, close to the $E_{2g}^1$ mode of the monolayer on Au, and a high-frequency peak matching that on SiO$_2$, suggesting that E(L) originates from the strained bottom layer, and the high-frequency peak originates from the top unstrained layer(s). (d) Comparison of the Raman spectra of 1-4L MoS$_2$ on SiO$_2$ and Au substrates. Dashed lines: Lorentzian fitting of doublets. Vertical line marks the peak position of the $E_{2g}^1$ mode of monolayer on Au. Scaling factors are normalized to the spectrum of monolayer on Au.

**ULF Raman: Shear and Breathing Modes**

Figure 3 shows the ULF Raman spectra of 1-10L $MoS_2$ on 10 nm Au and $SiO_2$ substrates. The weak interlayer interactions within 2D materials often manifest as ULF Raman modes with frequencies below 100 $cm^{-1}$.[23-26] Such modes normally consist of two types of vibrations: the *shear modes* for which individual layers move relative to each other parallel to the layer plane (Figure 3a) and the *breathing modes* for which the layers move perpendicularly to the layer plane (Figure 3b).

For $MoS_2$ on $SiO_2$, both the first-order shear modes (S1) and the first-order breathing modes (B1) are strong and clearly visible (Figure 3c). The S1 mode upshifts while the B1 mode downshifts with the number of layers, consistent with the literature.[23-25] The spectra on Au are quite different. The S1 modes on Au are clearly visible, but the B1 modes (marked by triangles) are significantly suppressed (Figure 3d). The exact mechanism for the suppression of the B1 modes is not clear. A similar phenomenon is attributed to the quasi-covalent bonding between $MoS_2$ and Au.[36] However, this factor alone cannot explain why the suppression is observed in crystals up to 10 layers and is selective for the breathing mode. We find that the intensity of the B1 modes varies with Au thickness and appears to be stronger on 2 nm and TS Au than on other Au substrates (Figure S3). This suggests that there are other mechanisms contributing to the quenching of the B1 modes, which will be a matter of future study.

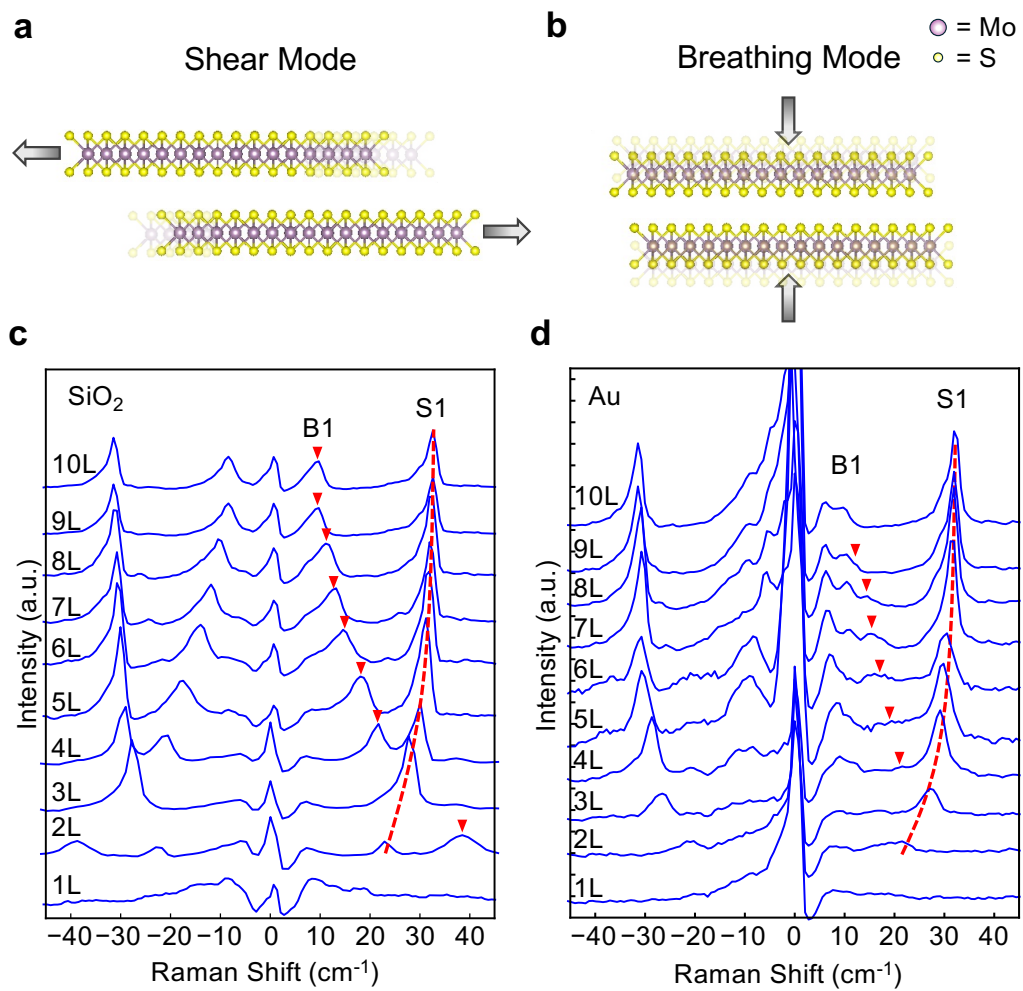

**Figure 3.** ULF Raman spectra of $MoS_2$. (a) and (b) Schematics of the vibrations of shear and breathing modes, respectively. (c) and (d) Measured ULF Raman spectra of 1-10L $MoS_2$ on $SiO_2$ (c) and 10 nm Au (d). The dashed lines serve as a guide to the eye for the first-order shear (S1) modes. Triangles mark the locations of the first-order breathing (B1) modes, which are notably suppressed on Au.

The S1 modes and B1 modes of $MoS_2$ on Au shift with respect to those on $SiO_2$. To see this more clearly, we plot the Stokes Raman spectra of 2-4L $MoS_2$ on Au and $SiO_2$ together in Figure 4a. The S1 modes on Au are downshifted with respect to those on $SiO_2$. Figure 4b presents the frequencies of S1 and B1 modes of 2-10L $MoS_2$ on Au and $SiO_2$. The S1 frequency of bilayer

MoS$_2$ on Au is downshifted by 2.2 cm$^{-1}$ compared to MoS$_2$ on SiO$_2$. The difference decreases with increasing MoS$_2$ thickness. By contrast, the B1 mode is upshifted for MoS$_2$ on Au compared to SiO$_2$.

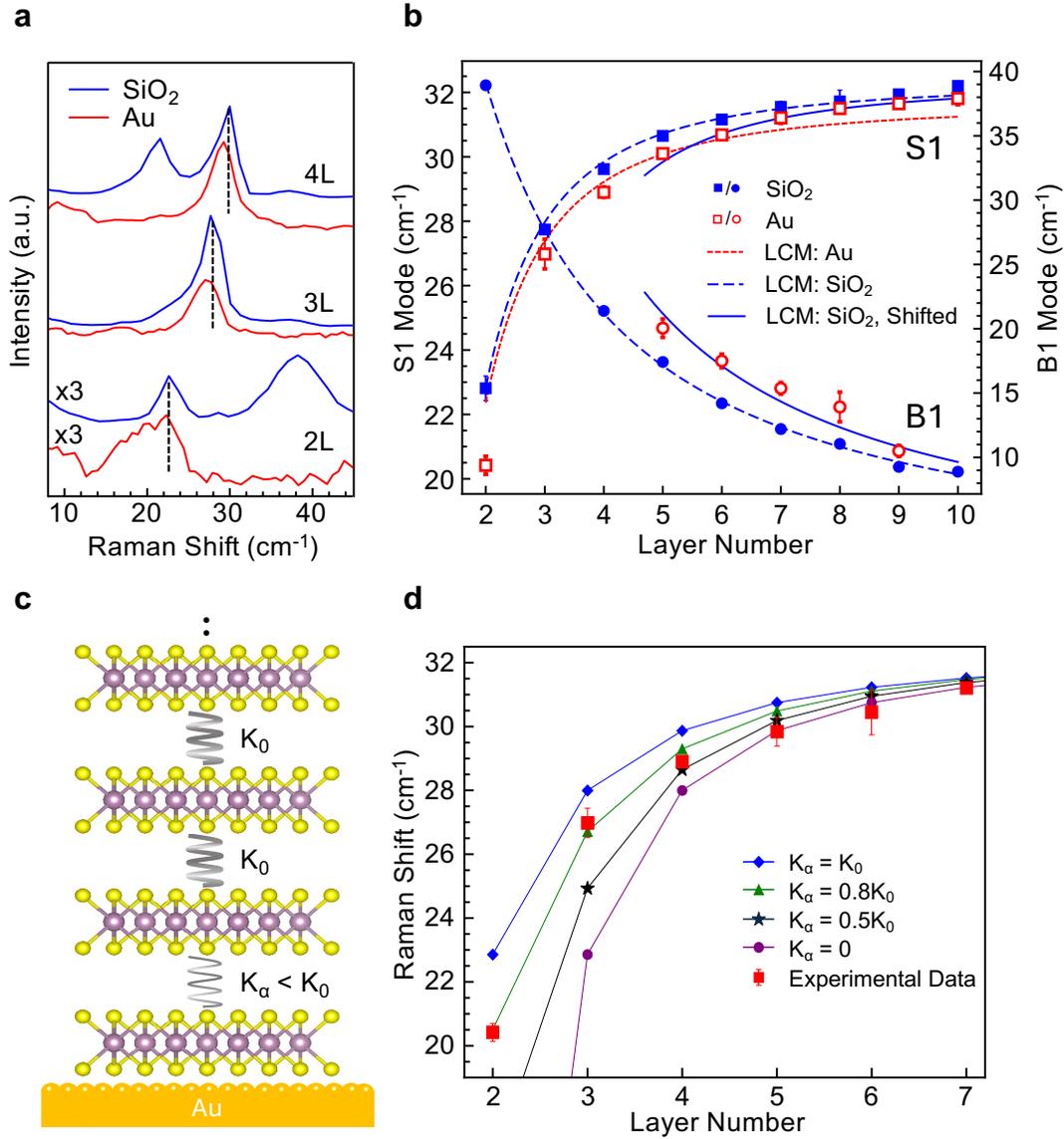

**Figure 4.** Layer-dependent ULF Raman modes of MoS$_2$ and LCM fitting. (a) ULF Stokes Raman spectra of 2-4L MoS$_2$ on SiO$_2$ and 10 nm Au substrates, showing the downshift of the S1 modes on Au with respect to those on SiO$_2$. Dotted lines mark the peaks of MoS$_2$ on SiO$_2$. (b) S1 and B1 modes of MoS$_2$ on SiO$_2$ (filled symbols) and 10 nm Au (unfilled symbols) substrates. Dashed blue

curves: LCM fitting of the S1 and B1 experimental data on SiO$_2$ substrates. Dotted red curve: LCM fitting of the S1 experimental data on Au substrates, which overestimates the frequencies of thin crystals (< 6 layers) and underestimates those of thicker ones (> 6 layers). Solid blue curves: dashed blue curves shifted to the right by one layer, well matching the data points of ≥ 5L MoS$_2$ on Au. (c) Schematic illustrating a modified linear chain model (mLCM) with a variable force constant $K_\alpha$ at the first MoS$_2$-MoS$_2$ interface. The force constants at other interfaces are kept at $K_0$, same as that of a pristine MoS$_2$. (d) Solid lines with symbols: S1 modes of MoS$_2$ calculated by mLCM of varying $K_\alpha$. Comparing the experimental data on Au with the calculated frequencies provides the reduced force constants $K_\alpha$ for different thickness layers. The $K_\alpha$ for bilayer and trilayer MoS$_2$ is about $0.8K_0$, for tetralayer it is about $0.5K_0$, and for crystals thicker than 5 layers it is nearly zero, indicating the bottom adhered layer is almost fully decoupled from the top layers.

**Linear Chain Model and Decoupling Effect**

To quantify these shifts, we employ a linear chain model (LCM), which can be used to calculate the shear and breathing modes of pristine 2D crystals. The model treats each individual layer as a mass point and the interaction between adjacent layers is represented by springs with a universal force constant. For a N-layer MoS$_2$ crystal, the model yields analytical solutions for the S1 and B1 mode frequencies, given by: [23,25]

$$\omega_{S1} = \sqrt{\frac{K_x}{2\mu\pi^2c^2}\left(1 - \cos\left(\pi\left(\frac{N-1}{N}\right)\right)\right)} \quad (1)$$

$$\omega_{B1} = \sqrt{\frac{K_z}{2\mu\pi^2c^2}\left(1 - \cos\left(\frac{\pi}{N}\right)\right)} \quad (2)$$

where $\omega_{S1/B1}$ are the thickness-dependent frequencies of the S1/B1 modes respectively, $K_{x/z}$ are the respective interlayer spring constants, $c$ is the speed of light and $\mu$ is the mass per unit area of MoS$_2$.

Both the S1 and B1 modes of MoS$_2$ on SiO$_2$ substrates can be well fitted with the LCM model (Figure 4b, filled symbols and dashed blue curves), obtaining a value of $(2.8 \pm 0.1) \times 10^{19}$ Nm$^{-3}$ for $K_x$ and a value of $(8.3 \pm 0.1) \times 10^{19}$ Nm$^{-3}$ for $K_z$. These agree well with the literature averages of $(2.8 \pm 0.1) \times 10^{19}$ Nm$^{-3}$ and $(8.8 \pm 0.1) \times 10^{19}$ Nm$^{-3}$ for $K_x$ and $K_z$ respectively.[24,25,37] The reliable fit by the LCM model indicates that SiO$_2$ substrates have minimal impact on the interlayer interaction of the exfoliated MoS$_2$ layers, confirming that SiO$_2$ is a good, weakly interacting reference substrate.

The results on Au, however, cannot be fitted with the LCM model with a universal force constant. Naïvely using this model to fit the data of MoS$_2$ on Au results in an overestimation of the frequencies of thin crystals ($N \leq 6$) while the frequencies of thicker ones ($N > 6$) are underestimated (dotted red line, Figure 4b). This implies that the force constants are not homogeneous across the interfaces of MoS$_2$ on Au. The observations that MoS$_2$ crystals preferentially cleave at the first interface and that the strain is mostly localized to the adhered layer, indicate that the force constant at the first interface must be different from those at other interfaces. In light of this, we employ a modified LCM (mLCM) to account for the S1 mode of MoS$_2$ on Au, where the force constant at the first MoS$_2$-MoS$_2$ interface, denoted as $K_\alpha$, is variable, while those between the remaining layers, $K_0$ (the force constant of pristine MoS$_2$), are kept constant, as schematically shown in Figure 4c. The Raman frequency for an N-layer system can then be extracted by solving for the eigenvalues of the corresponding $N \times N$ force constant matrix representing the equations of motion:[23]

$$\omega_i^2 \bar{u}_i = \frac{1}{2\pi^2 c^2 \mu} \bar{D}\bar{u}_i \quad (3)$$

where $\bar{D}$ is the force constant matrix considering only the nearest neighbor interaction and $\bar{u}_i$ is the $i^{th}$ phonon eigenvector with frequency $\omega_i$.

By calculating the frequencies of the S1 modes at varying $K_\alpha$ values and matching the calculated frequencies with the experimental data on Au, the $K_\alpha$ values of MoS$_2$ crystals of various thickness can be deduced. $K_\alpha$ is found to be thickness dependent, which is about 80% $K_0$ for bilayers, ~50% $K_0$ for tetralayers, and nearly 0% $K_0$ for crystals thicker than five layers (Figure 4d). A reduced $K_\alpha$ means a weakened coupling at the first interface. The striking finding is that $K_\alpha$ is nearly zero for crystals thicker than five layers. This suggests that the top layers are almost fully decoupled from the adhered bottom layer. In these circumstances, the system of an N-layer crystal is equivalent to a superposition of a monolayer attached to the Au substrate and a decoupled (N − 1) layer sitting on top. Under this assumption, the S1 and B1 modes of a N-layer (N ≥ 5) MoS$_2$ on Au should closely resemble that of a free-standing crystal of (N − 1) layers (since monolayer does not have the interlayer modes). To verify this, we shift the LCM fitting curves of the S1 and B1 modes of MoS$_2$ crystals on SiO$_2$ to the right by one layer (solid blue curves, Figure 4b) and find that they closely match the data points on Au for MoS$_2$ thicker than five layers. These phenomena are observed on all Au samples, including 2 nm, 5 nm, and commercially purchased 100 nm TS Au films (Supporting Figure S4), indicating that the observed decoupling effect is universal in GAE.

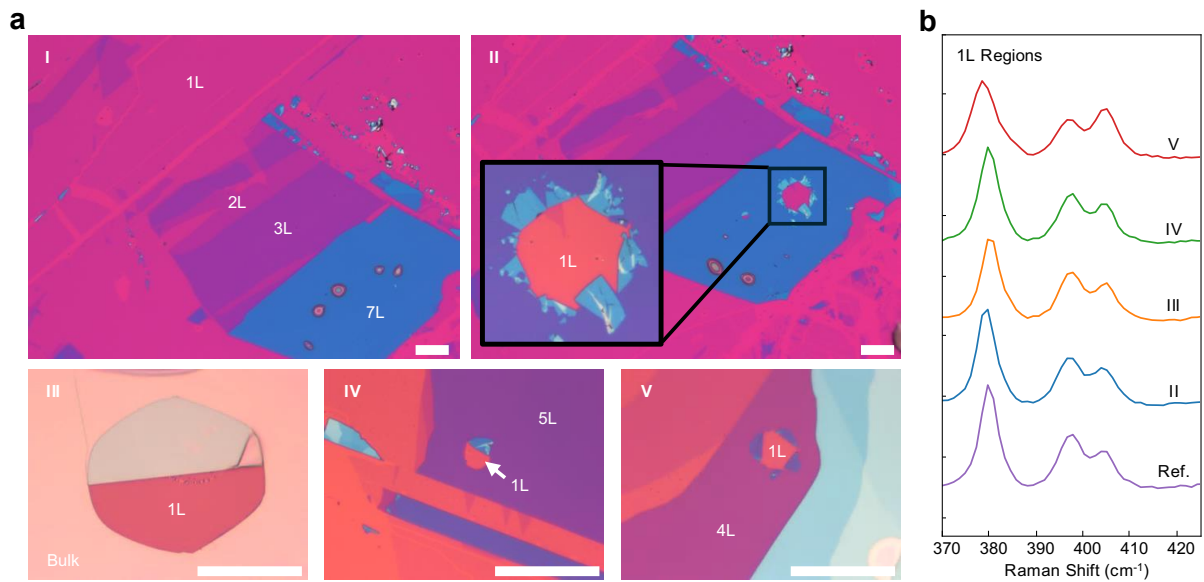

**Figure 5.** Bubbles in annealed MoS$_2$ on Au. (a) Optical images of MoS$_2$ on a 10 nm Au substrate after annealing at 200 °C, showing the formation of bubbles in thick layers (I), and after further annealing at 400 °C, showing the burst of bubbles and the exposed monolayers (II-V). (b) Raman spectra taken at the exposed regions (marked by '1L') in (a), confirming that the exposed MoS$_2$ are all monolayers. The reference spectrum of a monolayer MoS$_2$ on Au, measured on a non-heated sample, is included for comparison. All scale bars represent 25 μm.

**Bubbles: Direct Evidence of Decoupling**

When MoS$_2$ on 10 nm Au is annealed at 200 °C in ambient environment for 5 minutes, bubbles of sizes ranging from a few microns to tens of microns are observed to form in thick layers, but not in thin films (panel I, Figure 5a). When the samples are further annealed at 400 °C, the bubbles grow and some of them burst (panel II-V, Figure 5a). Interestingly, all the exposed surfaces are found to be monolayers (panel II-V, Figure 5a, supporting Figure S5), as confirmed by the measured Raman spectra, which show a single peak of E$_{2g}^1$ mode around 380 cm$^{-1}$ and splitting $A_{1g}$ modes, matching the spectrum of a reference monolayer on a non-heated Au sample (Figure 5b).

Nanoscale bubbles commonly exist within the weak interfaces of heterostructures or the interfaces between 2D materials and substrates, as the result of tiny air pockets trapped at locations of defects/contaminants. When heated, the air pockets move around and coalesce, forming micron-sized bubbles.[38,39] It is inevitable for our samples to contain some defects/contaminants (the samples are prepared in air), so it is normal to have some air pockets. Interestingly, we find micron-sized bubbles only forming in thicker flakes ( ≥4 layers). No microscopic bubbles are observed in 1-3L $MoS_2$ even after annealing at 400 °C (Figure 5a). This can be understood as follows. For thin flakes of two or three layers, although the interaction at the first $MoS_2$-$MoS_2$ interface is weakened, it is only by a small amount (about 20%), the layers are still quite strongly bonded, so the movement of trapped air pockets is restricted, unable to form large bubbles. It is even harder for bubbles to form underneath monolayers, as monolayers are strongly pinned on the Au substrate by quasi-covalent force. However, for crystals thicker than four layers, the first interface is significantly weakened, as discussed above. As such, when the samples are heated, air pockets can escape into the weakened interface, where they can move and coalesce, forming large bubbles (the top two small bubbles in the 7L flake of panel I merge into a large bubble and burst after further annealing, as shown in panel II, Figure 5a). Since the bubbles form at the weakened interface, when they burst a monolayer surface is exposed. These observations provide direct evidence unambiguously confirming that the first $MoS_2$-$MoS_2$ interface is the weakest link in the system and for thick crystals the adhered bottom layer is significantly decoupled from the top layers, in excellent agreement with the conclusions of the ULF Raman investigation.

**Mechanism of the Decoupling**

Strain is mentioned as the possible cause for weakened coupling at the first MoS$_2$ interface,[3,20,21] but these are only speculations without any solid experimental evidence or details of the weakening mechanism. To shed light on the effects of strain on the exfoliation, we conduct DFT simulations on the energy cost of exfoliation. For a N-layer crystal we compare the energetic cost of two configurations: (1) a system composed of one strained layer and a detached (N − 1) layer crystal free of strain; (2) the whole crystal is strained and allowed to relax. We define the exfoliation energy as $\Delta E_{exf} = \frac{E_{N-1}^0 + E_1}{A} - \frac{E_N}{A}$, where $E_{N-1}^0$ is the energy of (N − 1) unstrained layers, $E_1$ is the energy of a separate strained layer, $E_N$ is the energy of the whole strained stack that has been allowed to relax, and $A$ is the unit-cell area of the strained layer. Negative/positive values of $\Delta E_{exf}$ would indicate that monolayer exfoliation is energetically preferable/unfavorable.

The results for various strains and layer numbers of MoS$_2$ are presented in Figure 6. It shows for relatively small strains, $\Delta E_{exf}$ is positive, hence, exfoliation is unfavorable. For strains above 3.5%, $\Delta E_{exf}$ crosses the zero axis and becomes negative, indicating it is energetically favorable for the top stack to be detached from the bottom strained layer. The larger the strain, $\Delta E_{exf}$ crosses the zero axis at thinner layers and for crystals of same thickness, the exfoliation energy is negatively lower, suggesting it is more easily to exfoliate. The results reveal that a certain threshold strain is required to facilitate exfoliation and that the top stack tends to fully decouple from the strained bottom layer when it is thicker than a certain number of layers. Though the simulation is a simplified approximation of the actual experimental configuration, and the calculated threshold strain is larger than the strain measured experimentally, nevertheless the simulation results provide an insightful picture about the intertwined roles of strain and thickness of MoS$_2$ crystals, which are qualitatively in agreement with the experimental observations.

The simulation makes no explicit reference to any specific substrates, implying that the strain-induced decoupling effect is not limited to GAE, which also possibly plays a significant role in general mechanical exfoliations, including the scotch tape/$SiO_2$ method and metal-assisted exfoliation.

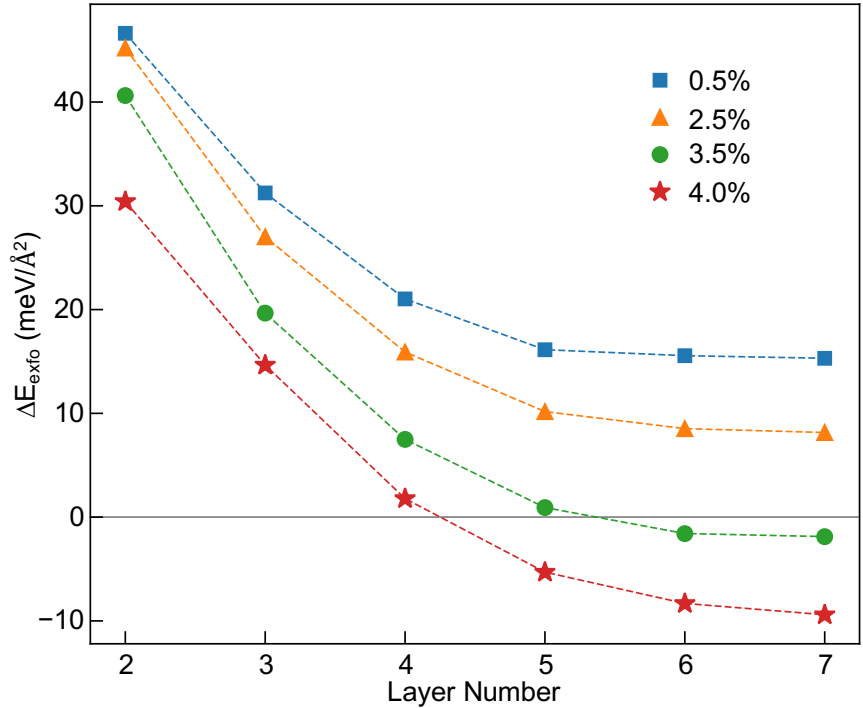

**Figure 6.** DFT simulation of the exfoliation energy of $MoS_2$ crystal. Calculated exfoliation energy and its variation with strain and number of layers. Negative/positive values of $\Delta E_{exf}$ indicate that monolayer exfoliation is energetically preferable/unfavorable.

As a final note, it is worthy to discuss the potential contribution of electrostatic force to the decoupling effect. Electrostatic repulsive force due to injected intercalation ions plays a significant role in electrochemical exfoliation of 2D materials.[40] However, in GAE, the impact of the electrostatic force due to the doped electrons by the Au substrate is minimal. As shown above, the concentration of doped electrons in the adhered layer is ~0.033 extra electron per unit cell.

Assuming the adjacent upper layer has the same concentration of doped electrons (which is an overestimate), the coulomb repulsive force between the adhered layer and the adjacent layer is $2.3 \times 10^{15}$ N per unit cell (the interlayer distance is taken as 0.65 nm), equivalent of a repulsive pressure of $2.5 \times 10^6$ Pa, which is only ~0.1% of the interlayer attractive pressure due to the vdW force in $MoS_2$ (2.4 GPa),[41] thereby the contribution of electrostatic force to the decoupling effect is negligible. The strain induced by the gold substrate is the primary cause for the decoupling effect in GAE.

CONCLUSIONS

In summary, using $MoS_2$ on Au as a model system, we establish that strain-induced decoupling is the primary mechanism of gold-assisted exfoliation (GAE). The gold substrate induces a biaxial strain in the adhered $MoS_2$ layer, which weakens the coupling at the first interface between the adhered layer and the adjacent layer. For $MoS_2$ crystals thicker than five layers, the coupling at the first interface is almost reduced to zero. This interface is the weakest link in the system. During exfoliation the 2D crystal preferentially cleaves at this junction, enabling the exfoliation of large-area monolayers. These findings provide valuable insights into the development of advanced exfoliation techniques and have broad implications in diverse research and technological domains.

Previously, it is thought that the adhesion to the substrate must exceed the interlayer vdW force within the 2D crystal to achieve the high-yield exfoliation.[1,2] The findings in this study suggest that this stringent requirement may not be necessary. Instead, the adhesion between the substrate and the 2D crystal needs only to surpass the weakened binding force at the first interface, which can be considerably weaker than the interlayer vdW force when strain is present. This indicates that even substrates with relatively weak adhesion to 2D materials can facilitate large-area exfoliation, provided that sufficient strain is induced in the adhered layer to effectively weaken the interfacial coupling. These findings highlight the potential of substrate engineering, particularly the enhancement of strain in the adhered layer, as a promising strategy to extend high-yield exfoliation techniques to a broader range of substrates and 2D materials, including dielectric substrates and non-chalcogen materials, beyond the scope of gold substrates.

The interface plays important roles in many physical and chemical phenomena (e.g., electrical and thermal conductivity, tunnelling effects, charge transfer, and Schottky barriers). As such, the discovery of significant weakening at the first interface of 2D crystals on gold substrates, which may also occur on other strain-inducing substrates, has broad implications across diverse disciplines and technologies, such as nanoelectrodes, field effect transistors, and photodiodes.

METHODS

**Sample Preparation**

All thin-film Au substrates are deposited on 100 nm $SiO_2$/Si substrates with a 1 nm Ti adhesion layer (except the samples of Figure 2, which are deposited on 300 nm $SiO_2$/Si substrates with a 3 nm Cr adhesion layer. The $SiO_2$/Si substrates and adhesion layer have no impact on exfoliation and the results). Prior to exfoliation, $SiO_2$/Si substrates are subsequently sonicated in acetone, isopropylalcohol, and deionized water. This is followed by plasma cleaning with a 25%$O_2$/75%Ar gas mixture or pure $O_2$. Reference samples are prepared by bringing an $MoS_2$ (HQ Graphene or Manchester Nanomaterials Ltd.) loaded tape into contact with the freshly cleaned and plasma activated $SiO_2$ surface. The samples used for ULF Raman are not heated to avoid introducing strain. For Au samples, bulk $MoS_2$ is first cleaved with a tape. The freshly cleaved crystal surface is then brought into contact with a freshly prepared Au film which is deposited onto the cleaned

SiO$_2$/Si substrate with an adhesion layer through magnetron sputtering. The Au (Birmingham Metal Ltd. or Micro to Nano Ltd.) and Ti (Testbourne Ltd. or Micro to Nano Ltd.) targets used were of >99.99% purity and the Cr (Micro to Nano Ltd.) is of > 99.95%. For all depositions the process chamber pressure is better than 9×10$^{-9}$ Torr. For template-stripped samples, pre-prepared chips of 100 nm Au(111) (Platypus Technologies) are peeled from the protective silicon wafer exposing a fresh clean Au surface onto which the crystal is exfoliated. In all cases exfoliation is carried out immediately after surfaces are exposed to air to avoid contamination.

**Raman Measurements**

ULF Raman measurements are carried out using a home-built ULF Raman apparatus operating at 532 nm in a backscattering configuration. The system is equipped with three BraggGate notch filters (BNF) allowing for ULF measurements down to 7 cm$^{-1}$. A 100× objective with N.A. of 0.9 is used for all measurements and the spot size is about 0.5 μm in diameter. Laser power is kept below 1.5 mW (measured at the sample) to avoid heating. Spectra are measured with an Acton SpectraPro SP-2300 spectrometer (Princeton Instruments) equipped with an iDus 416A CCD (Andor) and an 1800g/mm grating at 500 nm blaze. The spectral resolution of the system is 2.1 cm$^{-1}$ estimated from the full width at half maximum of the Rayleigh line and each pixel of the CCD spans 0.8 cm$^{-1}$. For analysis all peaks are fitted with a Lorentzian function and background subtraction is carried out as necessary. Positional error of peaks is taken to be the error of the fit. The high-frequency Raman spectra are obtained using a WITec alpha300 R Raman spectrometer (Oxford Instruments) with 532 nm excitation wavelength and 1800 g/mm grating. Laser is focussed through a 100× objective (Zeiss EC Epiplan-Neofluar) with power set to 0.5 mW to avoid damaging the samples.

**AFM Measurements**

AFM measurements are taken on an Asylum MFP-3D infinity machine in tapping mode with a 25 nm PPP-EFM tip (Nanosensors). Since the substrate interaction can introduce a large degree of variance in measured thicknesses, where possible the measurements of flakes are taken with respect to a monolayer to minimize the uncertainty.

**DFT Simulations**

The plane wave pseudopotential suite QUANTUM ESPRESSO is employed to perform fully self-consistent DFT-based electronic structure calculations by solving the standard Kohn-Sham (KS) equations.[42,43] Ultrasoft pseudopotentials from the PSlibrary are used for Mo and S atoms.[44] Kinetic-energy cut-offs are fixed to 80 Ry for electronic wave functions after performing rigorous convergence tests. The electronic exchange-correlation is treated under the generalized gradient approximation (GGA) that is parametrized by Perdew-Burke-Enzerhof (PBE) functional.[45] We adopt the Monkhorst-Pack scheme to sample the Brillouin zone in k-space with 12×12×1 grid.[46] Geometry optimization has been performed using the Broyden-Fletcher-Goldfrab-Shanno (BFGS) scheme.[47] The Convergence threshold of $10^{-10}$ and $10^{-5}$ are used on total energy (a.u) and forces (a.u) respectively for ionic minimization. The van der Waals interactions between layers are included into the calculation by incorporating DFT-D3 method.[48] We prepared the multi-layered $MoS_2$ systems with the software VESTA.[49]

ASSOCIATED CONTENT

**Supporting Information**

The Supporting Information is available free of charge.

The supporting information contains additional figures showing large area monolayers on all substrates used; AFM measurements carried out to confirm layer numbers; ULF Raman spectra of MoS$_2$ films exfoliated on various Au substrates; LCM and mLCM fits of the ULF Raman modes of MoS$_2$ on various Au substrates; additional images of bubbles.

## AUTHOR INFORMATION

**Corresponding Author**

*Jakob Ziewer, E-mail: jziewer01@qub.ac.uk

*Fumin Huang, E-mail: f.huang@qub.ac.uk

**Author Contributions**

The manuscript is written through contributions of all authors. All authors have given approval to the final version of the manuscript.

**Funding Sources**

UK EPSRC (EP/S023321/1), Czech Science Foundation (GA22-04408S), European Regional Development Fund, P JAC (project No. CZ.02.01.01/00/22 008/0004558).

**Notes**

The authors declare no competing financial interest.

## ACKNOWLEDGMENT

J. Z. thanks the UK EPSRC and SFI Centre for Doctoral Training in Photonic Integration and Advanced Data Storage (PIADS) program for the sponsorship of PhD studentship. M.H., L.P., and M.V. acknowledge the funding support by the Czech Science Foundation Project No. GA22-04408S. The support of European Regional Development Fund, P JAC (project No. CZ.02.01.01/00/22 008/0004558) is also acknowledged.